# A “kilonova” associated with short-duration $\gamma$-ray burst 130603B

N. R. Tanvir[1], A. J. Levan[2], A. S. Fruchter[3], J. Hjorth[4], R. A. Hounsell[3], K. Wiersema[1], R. L. Tunnicliffe[2]

[1]Department of Physics and Astronomy, University of Leicester, University Road, Leicester LE1 7RH, UK. [2]Department of Physics, University of Warwick, Coventry CV4 7AL, UK. [3]Space Telescope Science Institute, 3700 San Martin Drive, Baltimore, Maryland 21218, USA. [4]Dark Cosmology Centre, Niels Bohr Institute, University of Copenhagen, Juliane Maries Vej 30, 2100 Copenhagen, Denmark.

**Short-duration $\gamma$-ray bursts (SGRBs) are intense flashes of cosmic $\gamma$-rays, lasting less than ~2 s, whose origin is one of the great unsolved questions of astrophysics today[1,2]. While the favoured hypothesis for their production, a relativistic jet created by the merger of two compact stellar objects (specifically, two neutron stars, NS-NS, or a neutron star and a black hole, NS-BH), is supported by indirect evidence such as their host galaxy properties[3], unambiguous confirmation of the model is still lacking. Mergers of this kind are also expected to create significant quantities of neutron-rich radioactive species[4,5], whose decay should result in a faint transient in the days following the burst, a so-called “kilonova”[6-8]. Indeed, it is speculated that this mechanism may be the predominant source of stable *r*-process elements in the Universe[5,9]. Recent calculations suggest much of the kilonova energy should appear in the near-infrared (nIR) due to the high optical opacity created by these heavy *r*-process elements[10-13]. Here we report strong evidence for such an event accompanying SGRB 130603B. If this simplest interpretation of the data is correct, it provides (i) support for the compact object merger hypothesis of SGRBs, (ii) confirmation that such mergers are likely sites of significant *r*-process production and (iii) quite possibly an alternative, un-beamed electromagnetic signature of the most promising sources for direct detection of gravitational-waves.**

SGRBs have long been recognised as a distinct sub-population of GRBs[14]. If they are indeed produced by compact binary mergers, it would mean that SGRBs may provide a bright electromagnetic signal accompanying events detected by the next generation of gravitational-wave interferrometers[15]. Localising electromagnetic counterparts is an essential prerequisite to obtaining direct redshift measurements, and to further constraining the astrophysics of the sources. However, the evidence supporting this progenitor hypothesis is essentially circumstantial: principally that many SGRBs seem to reside in host galaxies, or regions within their hosts, lacking ongoing star formation, thus making a massive star origin unlikely (in contrast to long-duration bursts, which arise in the core-collapse of some short-lived massive stars[16]). Unfortunately, progress in studying SGRBs has been slow; Swift only localises a handful per year, and they are typically faint, with no optical afterglow or unambiguous host galaxy found in some cases despite rapid and deep searches.

SGRB 130603B was detected by the Burst-Alert-Telescope (BAT) on NASA's Swift satellite at 2013-06-03 15:49:14 UT[17], which measured its duration to be $T_{90} \approx 0.18 \pm 0.02$ s in the 15–350 keV band[18]. The burst was also detected independently by Konus-Wind which found a somewhat shorter duration, $T_{90} \approx 0.09$ s in the 18–1,160 keV band[19]. This places the burst unambiguously in the short-duration class, which is also supported by the absence of bright supernova emission which is generally found to accompany nearby long-duration bursts (see below). The optical afterglow was detected at the William Herschel Telescope[20], and found to overlie a galaxy previously detected in the Sloan Digital Sky Survey imaging of this field. The redshift of both afterglow[21] and host galaxy[22] were found to be $z = 0.356$.

Another proposed signature of a NS-NS/NS-BH binary merger is the production of a so-called "kilonova" (sometimes also termed a "macronova" or "*r*-process supernova") due to the decay of radioactive species produced and initially ejected during the merger process - in other words, an event similar to a faint, short-lived supernova[6-8]. Detailed calculations suggest that the spectra of such kilonova sources will be determined by the heavy *r*-process ions created in the neutron-rich material. Although these models[10-13] are still far from being fully realistic, a robust conclusion is that the optical flux will be greatly diminished by line-blanketing in the rapidly expanding ejecta, with the radiation emerging instead in the nIR, and stretched out over a longer time scale than would otherwise be the case. This makes previous limits on early optical kilonova emission unsurprising[23]. Specifically, the nIR light curves are expected to exhibit a broad peak, rising after a few days and lasting a week or more in the rest frame. The relatively modest redshift and intensive study of SGRB 130603B made it a prime candidate for searching for such a kilonova.

We imaged the location of the burst with the NASA/ESA Hubble Space Telescope (HST) at two epochs, the first ≈9 days post-burst, and the second at ≈30 days. On each occasion, a single orbit integration was obtained in both the optical F606W filter (0.6 μm) and the nIR F160W filter (1.6 μm) (full details of the imaging and photometric analysis discussed here are given in the Supplementary Information). The HST images are shown in Fig. 1; the key result is seen in the difference frames (right hand panels) that provide clear evidence for a compact transient source in the nIR in epoch 1 (we note that this source was also identified as a candidate kilonova in independent analysis of our epoch 1 data[24]), which has apparently disappeared by epoch 2 and is absent to the depth of the data in the optical.

At the position of the SGRB in the difference images, our photometric analysis gives $R_{606,\mathrm{AB}} > 28.25$ ($2\sigma$ upper limit) and $H_{160,\mathrm{AB}} = 25.73 \pm 0.20$. In both cases, we fitted a model point-spread function and estimated the errors from the variance of the flux at a large number of locations chosen to have similar background to that at the position of the SGRB. We note that some transient emission may remain in the second nIR epoch; experimenting with adding synthetic stars to the image leads us to conclude that any such late time emission is likely to be less than ~25% of the level in the first epoch in order for it not to appear visually as a faint point source in the second epoch, however, that would still allow the nIR magnitude in epoch 1 to be up to ~0.3 mag brighter.

In order to assess the significance of this result it is important to establish whether any emission seen in the first HST epoch could have a contribution from the SGRB afterglow. A compilation of optical and nIR photometry, gathered by a variety of ground-based telescopes in the few days following the burst, is plotted in Fig. 2, along with our HST results. Although initially bright, the optical afterglow light curve declines steeply after about ≈10 hr, requiring a post-break power-law decay rate of $\alpha \approx 2.7$ (where flux, $F \propto t^{-\alpha}$). The nIR flux, on the other hand, is significantly in excess of the same extrapolated power-law. This point is made most forcibly by considering the colour evolution of the transient which evolves from $R_{606} - H_{160} \approx 1.7 \pm 0.15$ at about 14 hr to greater than $R_{606} - H_{160} \approx 2.5$ at about 9 days. It would be very unusual, and in conflict with predictions of the standard external-shock theory[25], for such a large colour change to be a consequence of late-time afterglow behaviour. The most natural explanation is therefore that the HST transient source is largely due to kilonova emission, and in fact the brightness is well within the range of recent models over-plotted in Fig. 2, thus supporting the proposition that they are likely to be important sites of *r*-process element production. We note that this phenomenon is strikingly reminiscent, in a qualitative sense, of the red humps in the light curves of long-duration GRBs produced by underlying Type Ic SNe, although here the luminosity is considerably fainter, and the emission redder. The ubiquity and range of properties of the late-time red transient emission in SGRBs will undoubtedly be tested by future observations.

The next generation of gravitational wave detectors (Advanced-LIGO and Advanced-VIRGO) are expected to ultimately reach sensitivity levels allowing them to detect NS-NS and NS-BH inspirals out to distances of a few hundred Mpc ($z \approx 0.05$–$0.1$)[26]. However, no short-duration GRB has yet been definitely found at any redshift less than $z = 0.12$ over the 8.5 yr of the Swift mission to date[27]. This suggests that either the rate of compact binary mergers is worryingly low for gravitational-wave detection, or that most are not observed as bright SGRBs. The latter case could be understood if the beaming of SGRBs was rather narrow, for example, and hence the intrinsic event rate two or three orders of magnitude higher than that observed by Swift. Although the evidence constraining SGRB jet opening angles is limited at present[28] (indeed, the light curve break seen in SGRB 130603B may be further evidence for such beaming), it is clear that an alternative electromagnetic signature, particularly if approximately isotropic, such as kilonova emission, could be highly important in searching for gravitational-wave transient counterparts.

**Supplementary Information** is linked to the online version of the paper at www.nature.com/nature.

**Acknowledgements** This work was partly based on observations obtained at the Gemini Observatory, which is operated by the Association of Universities for Research in Astronomy, Inc., under a cooperative agreement with the US National Science Foundation on behalf of the Gemini partnership: the National Science Foundation (United States), the National Research Council (Canada), CONICYT (Chile), the Australian Research Council (Australia), the Ministério da Ciência e Tecnologia (Brazil) and SECYT (Argentina). This work was also partly based on observations made using ESO telescopes at the Paranal Observatory. The UKIRT is operated by the Joint Astronomy Centre on behalf of the UK Science and Technology Facilities Council. The UKIRT/WFCAM data used here were pipeline processed by the Cambridge Astronomical Survey Unit. The HST data were obtained under program GO/DD 13497. We thank the Space Telescope Science Institute director for approving and staff, particularly Denise Taylor, for expediting these Director's Discretionary Time observations. The Dark Cosmology Centre is funded by the Danish National Research Foundation. This work made use of data supplied by the UK Swift Science Data Centre at the University of Leicester. We acknowledge discussions with Antonio de Ugarte Postigo and Darach Watson.

**Author Contributions** N.R.T. lead the writing of the HST proposal, performed the final photometric analysis of the HST data and took primary responsibility for writing the text of the paper. A.J.L. contributed to all aspects of the observations and planning, particularly collating photometry and creating Fig. 2. A.S.F. and R.A.H. took primary responsibility for the detailed planning of the observations and the processing of the HST imaging. J.H., K.W. and R.L.T. contributed to planning the observing and analysis strategies. All authors contributed to the writing of the paper.

**Author Information** Reprints and permissions information is available at www.nature.com/reprints. The authors declare that they have no competing financial interests. Correspondence and requests for materials should be addressed to N.R.T. (nrt3@le.ac.uk).

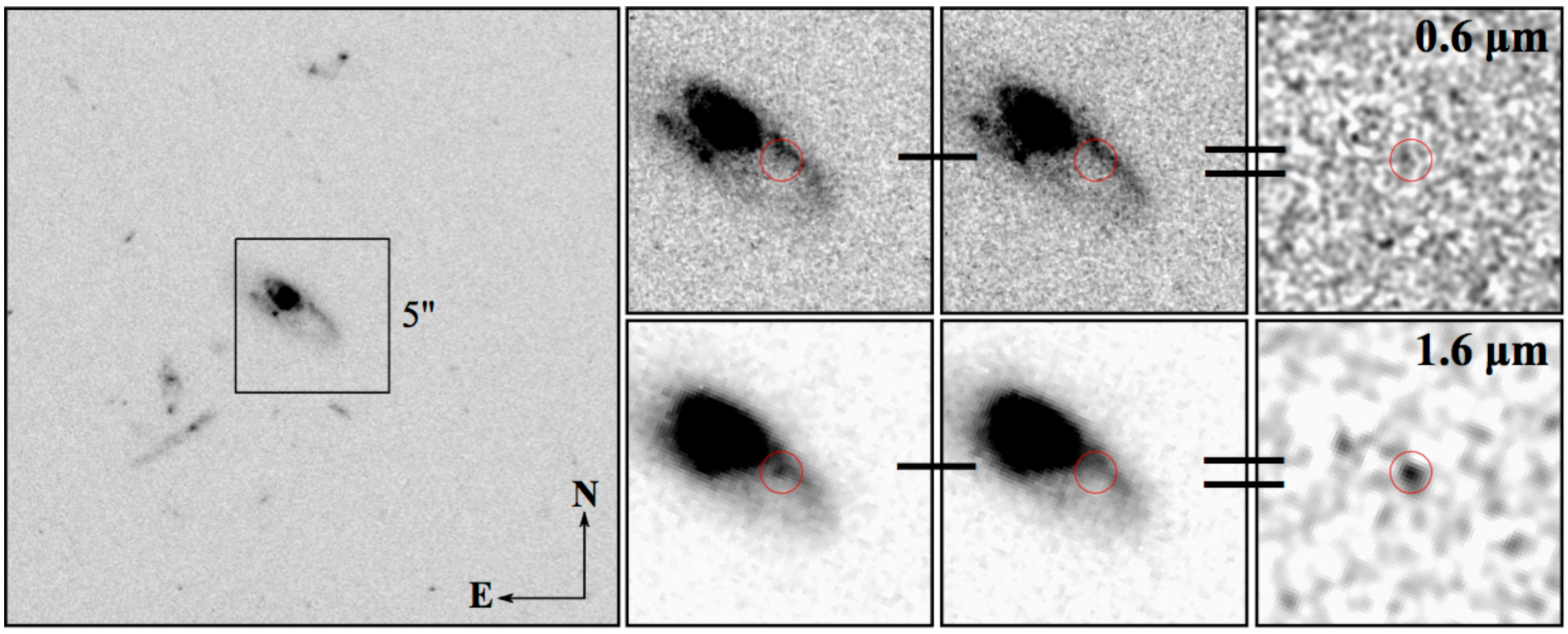

**Figure 1 HST imaging of the location of SGRB 130603B.** The host is well resolved and displays a disturbed, late-type morphology. The position (coordinates $RA_{J2000}$ = 11 28 48.16, $Dec_{J2000}$ = +17 04 18.2) at which the SGRB occurred (determined from ground-based imaging) is marked as a red circle, lying slightly off a tidally distorted spiral arm. The left-hand panel shows the host and surrounding field from the higher resolution optical image. The next panels show in sequence the first epoch and second epoch imaging, and difference (upper row F606W/optical and lower row F160W/nIR). The difference images have been smoothed with a Gaussian of width similar to the psf, to enhance any point-source emission. Although the resolution of the nIR image is inferior to the optical, we clearly detect a transient point source, which is absent in the optical.

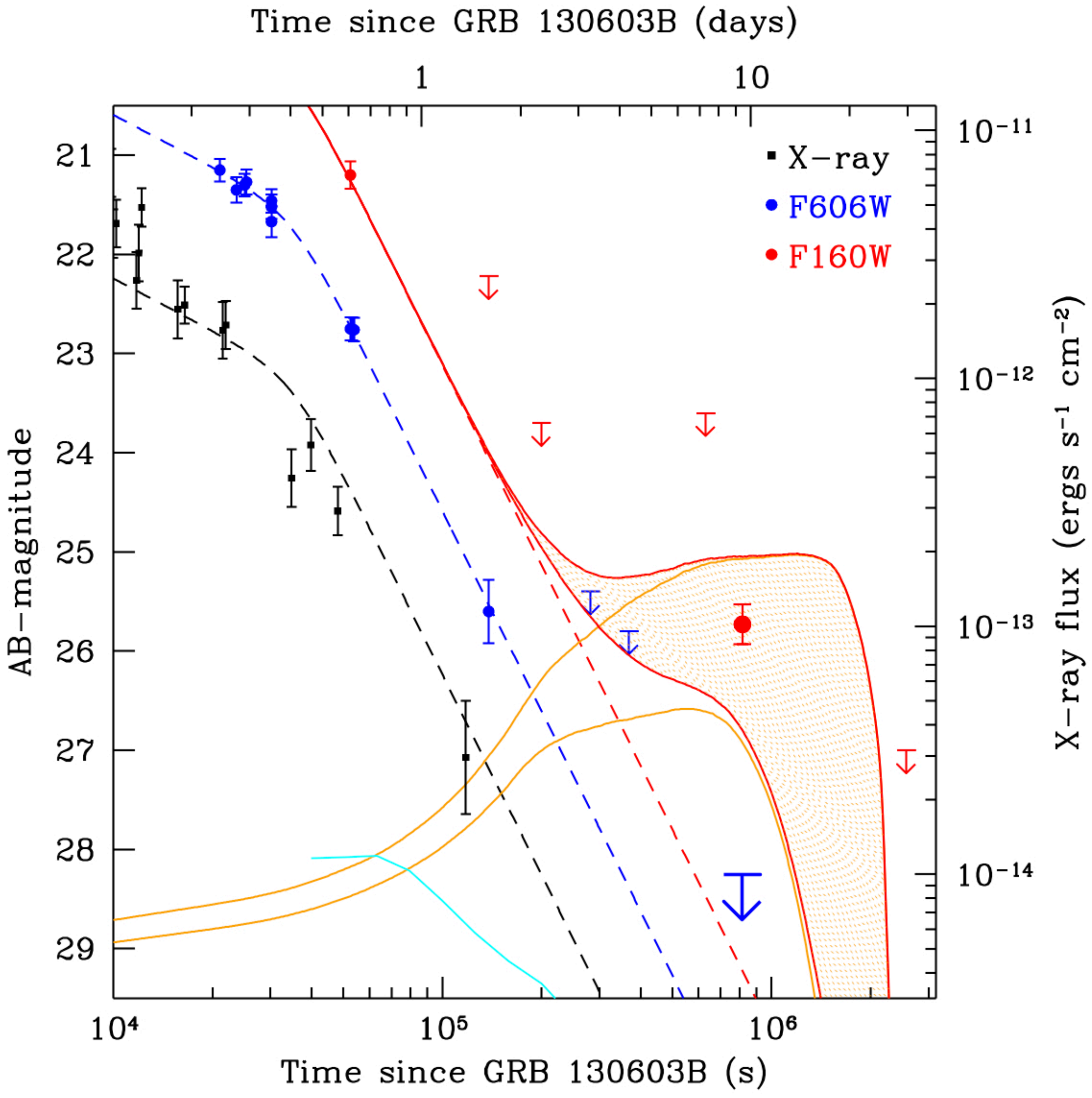

**Figure 2 Optical, near infrared (left axis) and X-ray (right axis) light curves of SGRB 130603B.** Upper limits are 2$\sigma$ and error bars 1$\sigma$. The optical data (gri bands) have been interpolated to the F606W band and the nIR data to the F160W band using an average spectral energy distribution at ≈0.6 days (see Supplementary Information). HST epoch 1 points are bold symbols. The optical afterglow decays steeply after the first ≈0.3 days, and is modelled here as a smoothly broken power-law (dashed blue line). We note that the complete absence of late-time optical emission also places a limit on any separate $^{56}$Ni driven decay component. The 0.3–10 keV X-ray data[29] are also consistent with breaking to a similarly steep decay (the dashed black line shows the optical light curve simply rescaled to match the X-ray points in this time frame), although the source dropped below Swift sensitivity by ~48 hr post-burst. The key conclusion from this plot

is that the source seen in the nIR requires an additional component above the extrapolation of the afterglow (red dashed line) assuming that it also decays at the same rate. This excess nIR flux corresponds to a source with absolute magnitude $M(J)_{AB} \approx -15.35$ at ~7 days post-burst in the rest frame. This is consistent with the favoured range of kilonova behaviour from recent calculations (despite their known significant uncertainties[11-13]), as illustrated by the model[11] lines (orange curves correspond to ejected masses of $10^{-2}$ $M_{\odot}$ [lower] and $10^{-1}$ $M_{\odot}$ [upper] respectively, and these are added to the afterglow decay curves to produce predictions for the total nIR emission shown as solid red curves). The cyan curve shows that even the brightest predictions for *r*-process kilonova optical emission are negligible.

# Supplementary Information

## 1. *HST* observations, processing and analysis.

A log of our *HST* observations is shown in Table 1. The standard STScI pipeline was used to initially process the data. The images from each epoch were then aligned to within a small fraction of a pixel using the task TWEAKREG from the STSDAS package DRIZZLEPAC[30] and combined using ASTRODRIZZLE. The final drizzled pixel scales were 15 pixels per arcsec for the WFC3/IR F160W images and 30 pixels per arcsec for the ACS F606W images.

To obtain photometry we adopted a very similar procedure for both optical and nIR data: (1) the point spread function (psf) was modelled with DAOPHOT[31] using several bright unsaturated stars in the epoch 1 image; (2) this psf was fitted to the transient source in the nIR difference image, and also at the same location in the optical difference image; (3) we adopted the standard *HST* zero-points to convert these fluxes in counts per second to AB magnitudes; (4) photometric errors were estimated in the case of the nIR observations by placing many artificial stars of the same magnitude as the transient at locations of the difference image corresponding to regions of similar background as provided by the host galaxy, and measuring the scatter in their recovered photometry. In the case of the optical photometry, since there was no significant flux at the transient location, we fitted psfs (allowing negative normalisations), again to locations of the difference image in regions similar to the host galaxy, and the error distribution was determined from the measured scatter in the results. This finally provided the following photometry: $R_{606,AB} > 28.25$ (2σ upper limit) and $H_{160,AB} = 25.73 \pm 0.20$.

## 2. Summary of ground monitoring

The early and mid-time photometry of the SGRB 130603B field as plotted in Figure 2 of the letter, is shown in Table 2. Further ground photometry and an expanded discussion, including its implications for the afterglow, will appear in de Ugarte Postigo *et al.* (in prep.).

These data, which include the original discovery observations of the optical afterglow, serve to monitor its behaviour over the first ~ 36 hours and place limits at later times. All data were de-biased and flat-fielded following standard procedures. The field lies within the SDSS footprint and so photometric calibration for griz observations is obtained directly from it. Near-IR calibration is taken from 2MASS, while we utilize our own calibration of the V-band.

To obtain photometric measurements of the afterglow in each band we performed image subtraction with the public ISIS code[32]. For clean subtractions we employed a later time image from each telescope as a template and subtract this from the earlier data (these were initially assumed afterglow free, but if an extrapolation of the power law decay suggested a low level of afterglow contamination, this was then reapplied as a small correction to the photometry of the subtracted image. These corrections were always less than 0.1 mag.). Photometric calibration of these subtracted images is obtained by the creation of an artificial star of known magnitude in the first image, with the errors estimated from the scatter in a large number of apertures (of radius approximately equal to the seeing) placed within the subtracted image. The placement of artificial stars close to the limiting magnitude within the image confirms that these can be recovered, and so the given limiting magnitudes are appropriate. However, we do note that the limiting magnitudes are based on the scatter in photometric apertures placed on the sky, not on the relatively bright regions of the host directly underlying the SGRB. Given that background errors are dominated by sky variance, this should not result in significant underestimate of errors: tests suggest an effect usually no more than 0.05 mag, so we conservatively add that extra error to all our magnitude uncertainties (in fact, we note that the scatter around the model fit did not suggest the original errors were underestimated, as mentioned below).

**Afterglow spectral energy distribution** Using this photometry we are able to construct a rather complete grizJHK spectral energy distribution (SED) at a time ≈ 0.6 days post-burst. We do not attempt to model this physically, but simply fit the data with a spline to allow interpolation. This SED is assumed to hold for the afterglow at later times, and the significantly increased slope at 9 days is one way of quantifying the evidence for an additional kilonova component.

## 3. Light curve analysis

In order to assess the significance of the late-time nIR enhancement, we performed the following analysis:

1. We fitted the optical data shown in Figure 2 of the paper (i.e. the optical detections interpolated to the R606,AB band using observed SED, and the *HST* epoch 1 F606W magnitude limit) with a smoothly broken power-law. The break time and smoothness were allowed to vary, as was the pre-break slope. The parameter of primary interest is the post-break slope, and for this we found a value $\alpha = 2.68$ (where flux $F \propto t^{-\alpha}$). We note that this fit has $\chi 2/\text{dof} = 5.1/7$, whereas if we had not added the estimate for an extra uncertainty due to image subtraction errors, the fit would have been essentially the same, but with $\chi 2/\text{dof} = 8.9/7$. In either case, the error estimates seem to be reasonable. On the low side the 95% confidence region goes to $\alpha = 2.18$, but on the high side it is not well constrained (i.e. steeper slopes become allowed by moving the break time later). We think it is unlikely that the late slope is much steeper than $\alpha \approx 2.7$, since that would not sit comfortably with the X-ray decay rate (see Figure 2 in the letter, and Ref. 33), although this makes no substantive difference to our conclusions. Note that if we just used imaging taken directly in an R-band (or r-band) filter, so avoiding significant interpolation, we still find a similar best fit slope $\alpha = 2.72$, and again the 95% confidence range only allows values as low as $\alpha = 2.18$.

2. We assumed the same light curve shape, including this late-time slope, applied to the H-band and hence determined the apparent magnitude of the excess nIR flux of $H_{160,AB} = 25.77^{+0.26}_{-0.22}$ ($1\sigma$ bounds). This value is not corrected for host dust extinction, but in the rest-frame J-band, this is unlikely to be more than 0.1–0.2 mag.

| UT date and start time | Exp (s) | Camera | Filter |
|---|---|---|---|
| 2013-06-12 23:11:55 | 2216.000 | ACS WFC | F606W |
| 2013-06-13 02:36:24 | 2611.751 | WFC3 IR | F160W |
| 2013-07-03 05:33:02 | 2611.751 | WFC3 IR | F160W |
| 2013-07-03 07:09:12 | 2216.000 | ACS WFC | F606W |

Table 1: Log of *HST* observations obtained with the Advanced Camera for Surveys/Wide Field Channel (ACS/WFC) and the Wide Field Camera 3/Infrared (WFC3/IR). Note, the original *Swift* trigger occurred at 2013-06-03 15:49:14 UT.

| MJD (start) | MJD (mid) | ΔT (days) | Telescope | Band | Exp. (s) | AB Mag |
|---|---|---|---|---|---|---|
| 56446.902986 | 56446.902986 | 0.244 | NOT/MOS | r | 5×360 | 21.15 ± 0.02 |
| 56446.923576 | 56446.932922 | 0.274 | WHT/ACAM | i | 3×300 | 20.86 ± 0.06 |
| 56446.948825 | 56446.949172 | 0.290 | GTC/OSIRIS | r | 30 | 21.30 ± 0.02 |
| 56446.943541 | 56446.952887 | 0.294 | WHT/ACAM | g | 3×300 | 21.90 ± 0.06 |
| 56446.988045 | 56446.988596 | 0.329 | FORS2 | V | 60 | 21.47 ± 0.02* |
| 56446.978 | 56446.989 | 0.330 | GMOS-S | g | 8×180 | 22.09 ± 0.04 |
| 56447.000 | 56447.011 | 0.352 | GMOS-S | r | 8×180 | 21.52 ± 0.05 |
| 56447.022 | 56447.032 | 0.373 | GMOS-S | i | 8×180 | 21.18 ± 0.11 |
| 56447.254670 | 56447.258471 | 0.599 | GMOS-N | z | 5×100 | 21.86 ± 0.03 |
| 56447.256481 | 56447.261765 | 0.603 | UKIRT | K | 70×10 | 21.06 ± 0.11 |
| 56447.262775 | 56447.266563 | 0.607 | GMOS-N | i | 5×100 | 22.26 ± 0.03 |
| 56447.267535 | 56447.272743 | 0.614 | UKIRT | J | 70×10 | 21.48 ± 0.14 |
| 56447.270813 | 56447.274604 | 0.615 | GMOS-N | r | 5×100 | 22.75 ± 0.03 |
| 56447.278897 | 56447.282685 | 0.623 | GMOS-N | g | 5×100 | 23.39 ± 0.04 |
| 56448.262940 | 56448.273513 | 1.61 | UKIRT | J | 140×10 | >22.5 |
| 56448.245230 | 56448.249620 | 1.59 | GMOS-N | g | 5×120 | >25.7 |
| 56448.254521 | 56448.259011 | 1.60 | GMOS-N | r | 5×120 | 25.6 ± 0.3 |
| 56448.264818 | 56448.269194 | 1.61 | GMOS-N | i | 5×120 | >24.7 |
| 56448.274065 | 56448.278411 | 1.62 | GMOS-N | z | 5×120 | >23.9 |
| 56448.965463 | 56448.976436 | 2.32 | HAWKI | J | 22×60 | >23.6 |
| 56449.914515 | 56449.918376 | 3.26 | GTC | r | 3×200 | >25.1 |
| 56450.919727 | 56450.923053 | 4.26 | GTC | r | 3×200 | >25.5 |
| 56453.950521 | 56453.961441 | 7.30 | HAWKI | J | 22×60 | >23.5 |

Table 2: Photometric observations of the SGRB 130603B afterglow, taken from our ob- servations, from de Ugarte Postigo *et al.,* and from Cucchiaria *et al.* 2013 (for Gemini-S; ref. 22). * Note that the V -band data is measured in a small (0.7 arcsec) aperture, but is not host subtracted. The errors given are statistical only and do not account for systematics between slightly different filter systems. The details of the subtraction also make small differences to the recovered flux, especially for sources sitting on moderately bright extended regions of their host galaxies. To account for this we estimate the additional variance on artificial stars inserted into the images to be $\sim$ 0.05 mag.

**References in SI**